\documentclass{PoS}
\usepackage[utf8]{inputenc}

\usepackage[table,x11names]{xcolor}
\usepackage{amssymb}	% Math
\usepackage{booktabs,amsmath} %table/tabular
\usepackage{bbm}
\usepackage{subcaption}
\usepackage{amsfonts}
\usepackage[skip=0.5\baselineskip]{caption}

\usepackage{geometry} 
\definecolor{boxcolor}{HTML}{ffe6a1}

\def \Ns {{N_{\sigma}}}
\def \Nt {{N_{\tau}}}
\def \Nc {{N_{c}}}

\def \hmu {{\hat{\mu}}}

\def \Zcal{\mathcal{Z}}

\newcommand{\lr}[1]{\left( #1 \right)}

\let\OLDthebibliography\thebibliography
\renewcommand\thebibliography[1]{
  \OLDthebibliography{#1}
  \setlength{\parskip}{0pt}
  \setlength{\itemsep}{0pt plus 0.3ex}
}

\title{Temporal Correlators in the Continuous Time Formulation of Strong Coupling Lattice QCD}

\ShortTitle{Temporal Correlators in CT limit of SC-QCD}

\author{\speaker{Marc Klegrewe}\\
Bielefeld University\\
        E-mail:  \email{mklegrewe@physik.uni-bielefeld.de}}
\author{Wolfgang Unger\\
Bielefeld University\\
E-mail: \email{wunger@physik.uni-bielefeld.de}}
\abstract{We present results for lattice QCD in the limit of infinite gauge coupling on a discrete spatial but continuous Euclidean time lattice. A worm type Monte Carlo algorithm is applied in order to sample two-point functions which gives access to the measurement of mesonic temporal correlators. The continuous time limit, based on sending $N_\tau\rightarrow \infty$ and the bare anistotropy to infinity while fixing the temperature in a non-perturbative setup, has various advantages: the algorithm is sign problem free, fast, and accumulates high statistics for correlation functions. Even though the measurement of temporal correlators requires the introduction of a binning in time direction, this discretization can be chosen to be by orders finer compared to discrete computations. For different spatial volumes, temporal correlators are measured at zero spatial momentum for a variety of mesonic operators. They are fitted to extract the pole masses and corresponding particles as a function of the temperature. We conclude discussing the possibility to extract transport coefficients from these correlators.}

\FullConference{The 36th Annual International Symposium on Lattice Field Theory - LATTICE2018\\
		22-28 July, 2018\\
		Michigan State University, East Lansing, Michigan, USA.}

\begin{document}

\section{Introduction}%uebliches bla bla mit Einleitung zu strong coupling partition function und anisotropy
The determination of the full QCD phase diagram, in particular the location of the critical point, is an important, long standing problem, requiring non-perturbative methods. In lattice QCD, several approaches have been developed to investigate the phase transition from the hadronic matter to the quark gluon plasma, but all of them are limited to small $\frac{\mu_B}{T}$ \cite{forcrand2009}. The reason for this is the notorious sign problem, which arises because the fermion determinant for finite  baryon chemical potential $\mu_B$ becomes complex, and importance sampling is no longer applicable. In QCD, the sign problem is severe. 
%The relative fluctuations of the complex phase factor grow exponentially with the lattice volume. Methods like complex Langevin or Lefshetz thimbles, which are based on complexifying parameter space, seem to be promising, however, they are not yet applicable to full QCD. 
Dual representations oftentimes solve or milden sign problems as being the case in strong coupling QCD (SC-QCD). 
%In the strong coupling limit of lattice QCD 
%In this limit, mesonic and baryonic degrees of freedom are resummed \cite{muetter1989} by reversing the order of integration: 
Here, first the gauge degrees of freedom are integrated out exactly, which allows replacing the Grassman integration by a sum over fermionic color singlets, resulting in a partition function being expressed as a gas of hadron world lines (c.f. monomer-dimer system \cite{{rossi1984}}). 
%There is no fermion determinant, and the sign problem is much milder. 
This representation allows us to obtain the full ($\mu_B, T$) phase diagram, and it shares important features of QCD such as confinement and spontaneous chiral symmetry breaking and its restoration at a transition temperature $T_c$ \cite{forcrand2010}. Moreover, the chiral limit can be studied very economically \--- simulations are faster than with a finite quark mass.
%Based on the 1-flavor (4 tastes) Euclidean continuous time formulation of SC-QCD, a Hamiltonian formulation can be derived \cite{unger2012} making it possible to perform simulations for arbitrary $N_f$ with a completely vanishing sign problem.
The Continuous Euclidean Time limit with its many assets (c.f. chap. 4) was first proposed to be applied to quantum field theories by  Beard and Wiese \cite{Beard1996}. Here, we use it to remove the sign problem completely.
\section{Strong Coupling QCD}
In SC-QCD, the gauge coupling is sent to infinity and hence the coefficient of the plaquette term $\beta=6/g^2$ is sent to zero. Thus, the Yang Mills part $F_{\mu\nu} F_{\mu\nu}$ is absent.
Subsequently, the gauge fields in the covariant derivative can be integrated out analytically. 
However, as a consequence of the SC-limit, the lattice spacing $a$  becomes very coarse, and no continuum limit can be achieved.
%The degrees of freedom in SC-QCD live on a crystal. 
We consider the SC-limit for staggered fermions.
The final partition function for the discrete system on a $\Ns^3\times \Nt$ lattice, after performing the Grassmann integrals analytically, is given by
\begin{align}
\Zcal(\gamma,N_\tau,m_q)=& \sum_{\{k,n,\ell\}}\prod_{b=(x,\hat{\mu})}\frac{(\Nc-k_b)!}{\Nc!k_b!}\color{blue}\gamma\color{black}^{2k_b\delta_{\hat{0},\hat{\mu}}}\prod_{x}\frac{\Nc!}{n_x!}(2am_q)^{n_x} \prod_l w(\ell,\mu) \label{SCPF}\\
\text{Grassmann constraint:} & \quad n_x+\sum_{\hat{\mu}=\pm\hat{0},\ldots \pm \hat{d}} \left( k_{\hat{\mu}}(x) + \frac{N_c}{2} |\ell_\mu(x)| \right)=\Nc, \quad 
\forall x\in \Ns^3\times \Nt \label{GC} \\
w(\ell,\mu)=\sigma(\ell)\color{blue}\gamma\color{black}^{\Nc \sum_x|\ell_0(x)|} & \exp\lr{\Nc \Nt r(\ell) a_\tau \mu}, \qquad \sigma(\ell)=(-1)^{r(\ell)+N_-(\ell)+1}\prod_{b=(x,\hmu)\in \ell}\eta_\hmu(x)\label{loops}
\end{align}
where \color{blue}$\gamma$\color{black}\, is the bare anisotropy coupling.
After this exact rewriting of the strong coupling partition function the system is described by 
confined, colorless, discrete degrees of freedom:
\begin{itemize}
\setlength\itemsep{0em}
\item Mesonic degrees of freedom $k_{\hat{\mu}}(x)\in \{0,\ldots \Nc\}$ (non-oriented meson hoppings called dimers) and
$n(x) \in \{0,\ldots Nc\}$ (mesonic sites called monomers).
\item Baryonic degrees of freedom, which form oriented baryon loops $\ell$ with sign $\sigma(\ell)=\pm 1$ and winding number $r(\ell)$ that depend on the geometry of the loops  Eq.~(\ref{loops}).
These loops are self-avoiding and do not touch the mesonic degrees of freedom.
\end{itemize}
Both mesonic and baryonic degrees of freedom obey the Grassmann constraint Eq.~(\ref{GC}). 
%Such a constrained configuration of dimers and baryon loops is presented in Fig.~(\ref{compareConfigs}). 
Monomers are absent since we will restrict to the chiral limit $m_q=0$. %($\prod_x \frac{N_c!}{n_x!}\left( 2am_q \right)^{n_x}=1$) 

\section{Anisotropic lattices}
%The weights for temporal meson or baryon hoppings in Eq.~(\ref{SCPF}) and Eq.~(\ref{loops}) contain the highlighted anisotropy parameter \color{blue}$\gamma$\color{black}. 
On a bipartite lattice with staggered fermions an even number of lattice points is required in all directions. Thus, the highest temperature that is possible to be addressed on an isotropic lattice is $aT=1/N_\tau=0.5$, with $N_
\tau$ the temporal extent. This is significantly too low to reach the critical temperature of chiral restoration. So, in practice, anisotropic lattices are chosen to study thermodynamical properties of staggered lattice QCD, in particular across the chiral phase transition. 
%This is why the highlighted bare anisotropy parameter \color{blue}$\gamma$\color{black} is present in the weights for temporal meson or baryon hoppings of Eq.~(\ref{SCPF}) and Eq.~(\ref{loops}). 
The anisotropy parameter $\xi=\frac{a}{a_\tau}$ is introduced into the definition of the lattice temperature
\begin{equation}
 T=\frac{1}{a_\tau N_\tau}=\frac{\xi(\gamma)}{a N_\tau}.
 \label{latticeTemp}
\end{equation}
which allows to assign different extents in spatial and temporal direction and thus, to modify the temperature continuously even above the chiral transition. $\xi$ becomes unity when the lattice is isotropic and diverges in the CT limit $a_\tau\rightarrow 0$.
%Now, the functional dependence of the ratio $\xi$ on $\gamma$ is of great interest and estimated in the weak coupling limit to be ($\xi(\gamma)=\gamma$) or by mean field theory based on a $1/d$-expansion \cite{Bilic1992a} as $\xi(\gamma)=\gamma^2$. 
As highlighted in Eq.~(\ref{latticeTemp}) the anisotropy parameter depends on the bare anisotropy coupling $\gamma$. However, the exact functional correspondence is unknown. Recent non-perturbative studies \cite{unger2017} suggest that
%However, the most reasonable approach describing the non-monotonic behavior is given by non-perturbative studies \cite{unger2017} that suggest 
\begin{equation}
\xi(\gamma)\simeq \kappa\gamma^2 + \frac{\gamma^2}{1+\lambda\gamma^4}, \quad \kappa=0.781(1) \;\;\text{for}\;\;\text{SU}(3).
\end{equation}
Further simplification is achieved by eliminating $\gamma$ and $N_\tau$, and to replace them by the temperature $aT$ completely. This is summarized in the continuum limit in Euclidean time:
\begin{equation}
\Nt\rightarrow \infty, \qquad \gamma \rightarrow \infty, \qquad \kappa\gamma^2/\Nt\equiv aT \;\; {\rm fixed.}
\end{equation}
Here, $\kappa\gamma^2/\Nt$ represents the temperature $aT$ in a well defined setup. Only one parameter is left that sets the thermal properties, and all discretization errors introduced by a finite $\Nt$ are removed.
\section{Continuous Time Limit and worm algorithm}%Herleitung partition function and properties
Designing an algorithm that operates in the continuous time (CT) limit will have several advantages: Since there is no need to perform the continuum extrapolation $\Nt \rightarrow \infty$, critical temperatures can be estimated more precisely, with a faster algorithm which only depends on one parameter, the temperature T.  Moreover, ambiguities arising from the functional dependence of observables on the anisotropy parameter will be circumvented.
Also in the baryonic part of the partition function great simplifications occur: Baryons become static in the CT limit for $N_c\geq 3$, hence, the sign problem is completely absent.
The CT partition function is obtained by the joined limit $\gamma$ and $N_\tau\rightarrow \infty$ and includes:
\begin{itemize}
\setlength\itemsep{-0em}
\item dimer contributions of Eq.~(\ref{SCPF}) are factorized into a spatial and temporal part and $\mathcal{Z}(\gamma,N_\tau)$ is rewritten such that spatial dimers obtain a weight $\gamma^{-2}$ \cite{unger2011}.
\item the limit $\gamma \rightarrow \infty$ implies that configurations with only zero or single spatial dimers contribute while configurations with multiple spatial dimers are considered to be suppressed.
\item configurations are characterized fully by dimers in the zero time slice $k_0(0)\in \{0,\dots N_c\}$ and bonds occupied by single spatial dimers which form vertices. Intervals between vertices have a weight of one and are omitted.
%Only locations where spatial and temporal dimers form vertices have an impact on the weight of a configuration
\item the limit $N_\tau\rightarrow \infty$ removes lattice artifacts in $a_\tau$ completely. Due to the even-odd decomposition there are $\frac{N_\tau}{2}$ positions available to distribute an oriented spatial dimer which gives rise to the factor $\frac{1}{2aT}$ in Eq.~(\ref{ZCT}).
\end{itemize}
Finally, a merely T dependent partition function is obtained
\begin{align}
\begin{split}
(N_c=3): \quad \quad \quad \mathcal{Z}(T)&=\sum_{k\in 2\mathbb{N}}\left( \frac{1}{2aT} \right)^k\sum_{\mathcal{G}'\in\Gamma_k}e^{\mu_B B/T}\hat{\nu}^{N_\bot}_\bot \quad \text{with} \;\;  \hat{\nu}^{N_\bot}_\bot=2/\sqrt{3} \\
\text{and} \;\; k&=\sum_{b=(x,\hat{i})}k_b=\frac{N_\llcorner+N_\bot}{2}, \quad N_{\llcorner/\bot}=\sum_x n_{\llcorner/\bot}(x)
\label{ZCT}
\end{split}
\end{align}
with the baryon number $B$, a non-trivial vertex weight $\hat{\nu}^{N_\bot}_\bot$, the number of $L/T-$shaped vertices $N_\llcorner/N_\bot$ and $\Gamma_k$ being the set of equivalence classes $\mathcal{G}'$ of graphs containing a total number $k$ of spatial hoppings, equivalent up to time shifts of the vertices.

Now, to sample this CT partition function a worm type algorithm is used, similar to the directed path algorithm introduced for SC-QCD in \cite{adams2003}. %The updating rules are outlined in Fig.~\ref{absorptionemission}. 
In analogy to the decomposition of the lattice into active and passive sites, we decompose the lattice into emission and absorption sites. By definition the worm tail is located at an absorption site and violates Eq.~(\ref{GC}). As a consequence, the worm head propagates through the lattice and restores (violates) the constraint in turns while visiting emission (absorption) sites respectively. 
During propagation the worm head either stops at an absorption site connected to a spatial dimer or emits a spatial dimer after some distance $\Delta \beta$ established by a Poisson process. The Poisson process assures that the oriented vertices, which always connect an emission and an absorption site, are exponentially distributed
\begin{align*}
\begin{split}
P(\Delta\beta)=\exp(-\lambda \Delta\beta),\quad \Delta\beta \in [0,\beta=1/aT], \quad
\lambda=d_M(x,t)/4, \quad d_M(x,t)=2d-\sum_{\hmu} n_B(x+\hmu)
\end{split}
\end{align*}
with $\lambda$ the ``decay constant'' for spatial dimer emissions. 
Due to the presence of baryons, $\lambda$ is space-time dependent, with $d_M(x,t)$ being the number of mesonic neighbors at a given coordinate. %Note that worm estimators, observables that are measured while worm propagation, are only accumulated when the head visits an absorption site. 
Throughout the worm evolution monomer-monomer two-point correlation functions are accumulated whenever the Grassmann constraint is restored by taking into account the respective positions of worm tail and head:
\begin{equation}
C(t_H-t_T,\vec{x}_H-\vec{x}_T)=C(\tau,\vec{x})=N_c\frac{O(C(\tau,\vec{x}))}{\text{\#worm updates}}.
\end{equation}
Such worm estimators are incremented as $O(C(\tau,\vec{x}))\rightarrow O(C(\tau,\vec{x}))+f(\dots)\cdot \delta_{x_T,x_1}\delta_{x_H,x_2}$ with
\begin{align}
\begin{split}
\text{discrete time:} \; &f(\gamma), \;\;\; \tau \in[0, \dots N_\tau] \\
\text{continuous time:} \; &f(T),\;\; \tau \in[0, \dots 1/T].
\label{increments}
\end{split}
\end{align}
By summing over them yields immediately the chiral susceptibility:
\begin{equation}
\chi_{\sigma,DT}=\frac{1}{V}\sum_{\vec{x},\tau} C(\tau,\vec{x}) \quad \text{and} \quad \chi_{\sigma,CT}=\frac{1}{V}\sum_{\vec{x}}\int_{0}^{1/T}d\tau \;C(\tau,\vec{x}).
\end{equation}
\section{Temporal Correlators}
As for temporal correlators in CT, the increment $f(T)$ of Eq.~(\ref{increments}) is spread out to bins across the path covered by the worm head in temporal direction. Thus, even for the CT algorithm a discretization is introduced, however, it can be chosen by orders finer in comparison to a discrete time lattice extent. It is distinguished between two different histograms, either with even (absorption-absorption) or odd (absorption-emission) temporal distance contributions. Combinations of these histograms allow to construct correlators for the Non-Oscillating (NO) and Oscillating (O) channel. Additionally, by including the sign $g^D_x$ listed in Table \ref{signTable} various states are addressed for $N_f=1$.
Since temporal correlators at zero spatial momentum are measured, the extracted meson masses are pole masses ($E_0(\vec{p}=0)=m_0$). The respective correlators are expressed as a sum over the staggered fermion fields $\bar{\chi}_x\chi_x$ with a diagonal dirac-taste kernel ($\Gamma^D=\Gamma^{F*}$), that is realized by the signs $g^D_x$:
\begin{equation}
C(t)=\sum_{\vec{x}}\langle \bar{\chi_0}\chi_0\bar{\chi}_{\vec{x},t}\chi_{\vec{x},t}\rangle\cdot g^D_x.
\label{signGD}
\end{equation}
%%%%%%%%%%%%%%%%%%%%%%%%%%%%%%%%%%%%%%%%%%%%%%%%%%
\begin{table}[]
\caption{The sign $g_x^D$ defined in Eq.~(\ref{signGD}) yields different correlators for the kernel ($\Gamma^D \otimes \Gamma^F , \Gamma^D = \Gamma^{F*}$). Corresponding continuum and particle states for $N_f$=1 are named.}
\centering
\begin{tabular}{c | c c | c c | c c  }
\toprule
$g_x^D$ & \multicolumn{2}{c|}{$\Gamma^D\otimes\Gamma^F$} & \multicolumn{2}{c|}{$J^{PC}$} & \multicolumn{2}{c}{Physical states} \\
& NO & O & NO & O & NO & O \\
\midrule
$1$ 		  				&	$1\otimes 1$																	& $\gamma_0\gamma_5\otimes(\gamma_0\gamma_5)^*$ 	& $0^{++}$ 	& $0^{-+}$ 	& $\sigma_S$ 	& $\pi_A$		\\
$(-1)^{x_i}$   		&	$\gamma_i\gamma_5\otimes(\gamma_i\gamma_5)^* $ 	& $\gamma_i\gamma_0\otimes(\gamma_i\gamma_0)^*$	& $1^{++}$ 	& $1^{--}$ 	& $a_A$  		& $\rho_T$	\\
$(-1)^{x_j+x_k}$	&	$\gamma_j\gamma_k\otimes(\gamma_j\gamma_k)^*$	& $\gamma_i\otimes\gamma_i^*$  									& $1^{+-}$ 	& $1^{--}$ 	& $b_T$  		& $\rho_V$	 \\
$(-1)^{x_i+x_j+x_k}$		&	$\gamma_0\otimes\gamma_0^*$ 									& $\gamma_5\otimes(\gamma_5)^*$ 								& $0^{+-}$ 	& $0^{-+}$ 	& $-_V$  			& $\pi_{PS}$	\\
\midrule
\end{tabular}
\label{signTable}
\end{table}
%%%%%%%%%%%%%%%%%%%%%%%%%%%%%%%%%%%%%%%%%%%%%%%%%%
Hereafter, the workflow for continuous time pole mass extraction as well as discrete time is highlighted and results are compared. In order to obtain discrete time temporal correlators the even and odd histograms are fitted via a four parameter ansatz respectively:
\begin{align}
\begin{split}
C_{\text{DT,Even}}(\tau)=a_{NO}\cosh(m_{NO}(\tau-N_\tau/2))&\color{red}{-}\color{black}a_O\cosh(m_O(\tau-N_\tau/2) \\
C_{\text{DT,Odd}}(\tau)=\underbrace{a_{NO}\cosh(m_{NO}(\tau-N_\tau/2))}_{\text{Non-oscillating Correlator}}&\color{red}{+}\color{black}\underbrace{a_O\cosh(m_O(\tau-N_\tau/2)}_{\text{Oscillating Correlator}}.
\end{split}
\end{align}
A combined fit to simultaneously describe both data sets is possible, but more challenging when it comes to fit convergence.
Finally, by addition/subtraction the correlators are as follows:
\begin{align}
\begin{split}
C_{\text{DT,NO}}(\tau)=\frac{1}{2}\left(C_{\text{DT,Even}}(\tau)\color{red}+\color{black}C_{\text{DT,Odd}}(\tau)\right), \quad
C_{\text{DT,O}}(\tau)   =\frac{1}{2}\left(C_{\text{DT,Even}}(\tau)\color{red}-\color{black}C_{\text{DT,Odd}}(\tau)\right).
\end{split}
\end{align}
\begin{figure}
\centering
\begin{subfigure}[c]{0.45\textwidth}
\includegraphics[width=1.00\textwidth]{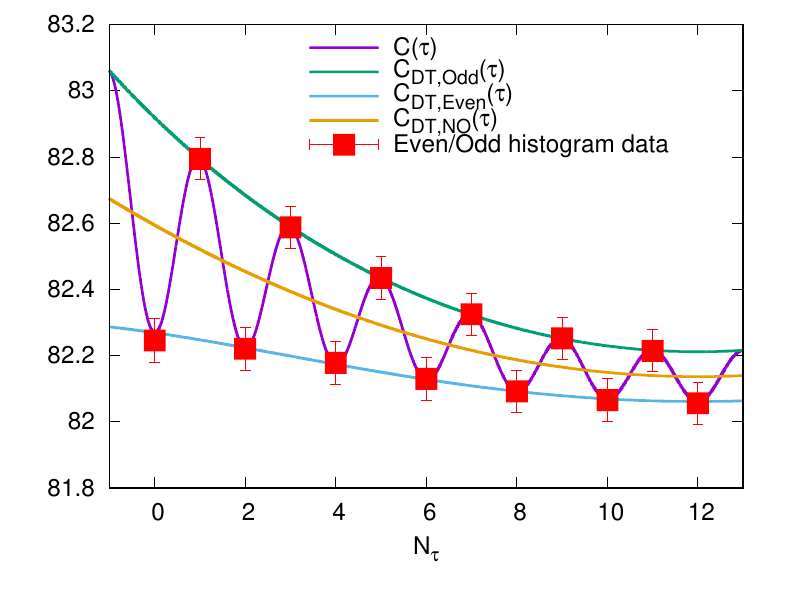}
\subcaption{Histogram data fitted by several approaches, leading to the constructed correlator $C_{DT,NO}(\tau)$.}
\label{dtCorrFit}
\end{subfigure}
\begin{subfigure}[c]{0.04\textwidth}
\phantom{....}
\end{subfigure}
\begin{subfigure}[c]{0.45\textwidth}
\includegraphics[width=1.00\textwidth]{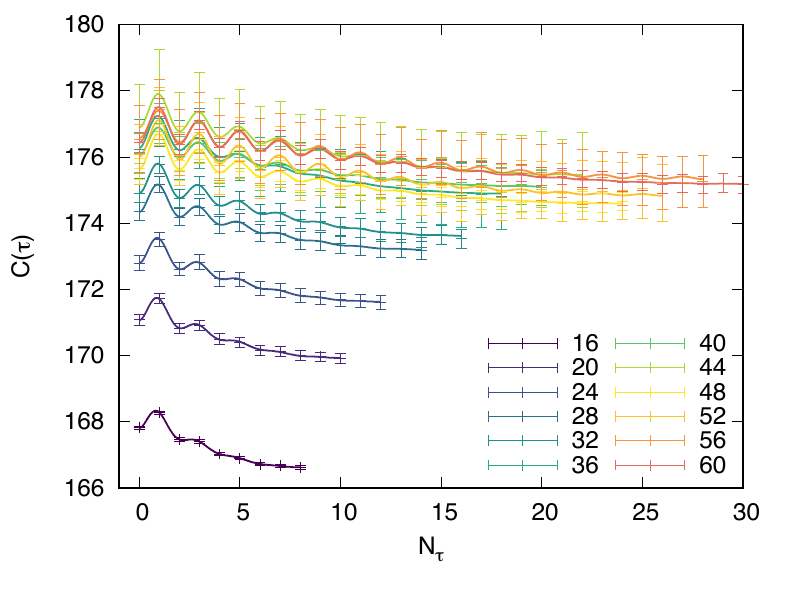}
\subcaption{Histogram data fits for multiple temporal extension ranging from $N_\tau=16$ to $N_\tau=60$.}
\label{dtCorrFitVar}
\end{subfigure}
\begin{subfigure}[c]{0.45\textwidth}
\includegraphics[width=1.00\textwidth]{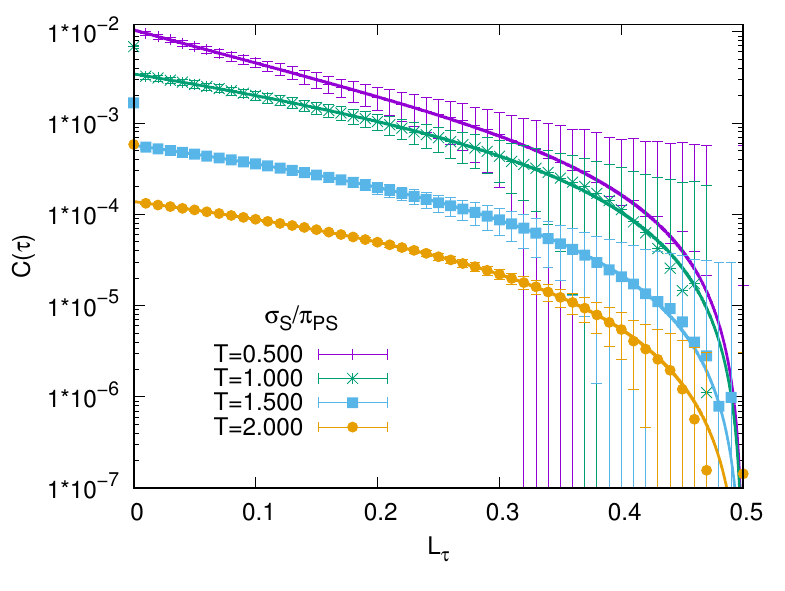}
\subcaption{Continuous time histogram data fitted according to Eq.~(\ref{fitCT}) for the mass degenerated channels $\sigma_S/\pi_{PS}$.}
\label{compareMasses}
\end{subfigure}
\begin{subfigure}[c]{0.04\textwidth}
\phantom{....}
\end{subfigure}
\begin{subfigure}[c]{0.45\textwidth}
\includegraphics[width=1.00\textwidth]{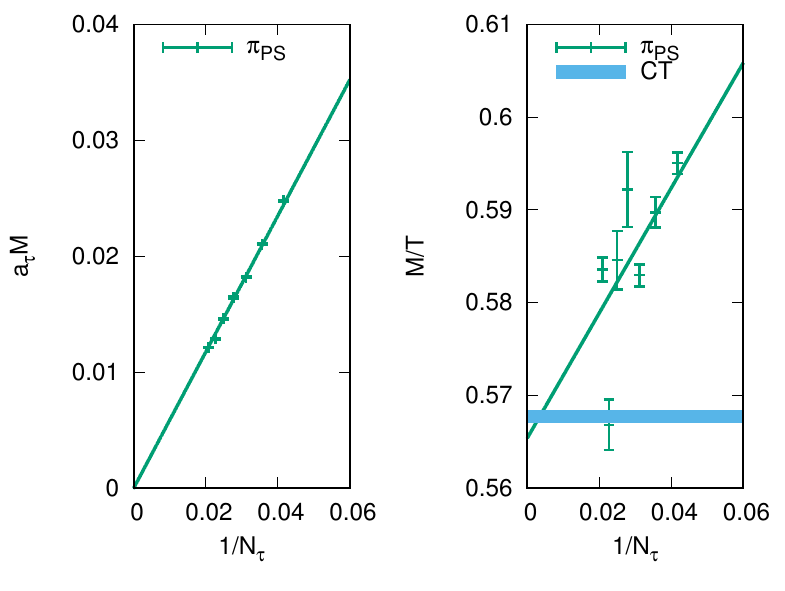}
\subcaption{Extrapolated masses from discrete time simulations. Rescale to units of $M/T$ and compare with continuous time simulation result (blue band).}
\label{dtExtrap}
\end{subfigure}
\end{figure}
Note that in discrete time only histogram data sets are described by the fits as presented in Fig.~(\subref{dtCorrFit}), however, the final constructed correlators are not. Now, these fits have to be performed for various $N_\tau$ (c.f. Fig.~(\subref{dtCorrFitVar})) such that an appropriate $N_\tau \rightarrow \infty$ extrapolation can be carried out (c.f. Fig.~(\subref{dtExtrap})). Finally, this workflow is necessary for the different channels and multiple temperatures. 
In comparison, the added and subtracted histograms out of continuous time simulations
\begin{align}
\begin{split}
C_{\text{CT,O}}(\tau)=a_{NO}\cosh(m_{NO}(\tau-1/2))=&\frac{1}{2}(C_{\text{Odd
}}(\tau)+C_{\text{Even}}(\tau)) \\
C_{\text{CT,NO}}(\tau)=a_O\cosh(m_O(\tau-1/2))=&\frac{1}{2}(C_{\text{Odd}}(\tau)-C_{\text{Even}}(\tau)) 
\label{fitCT}
\end{split}
\end{align}
give directly rise to the (Non)-Oscillating correlators respectively and are fitted in accordance with Eq.~(\ref{fitCT}) as shown in Fig.~(\subref{compareMasses}). The resulting pole masses are measured in $M/T$.
% such that the discrete time results have to be rescaled (c.f. Fig.~(\subref{dtExtrap})).
\begin{figure}
\centering
\begin{subfigure}[c]{0.45\textwidth}
\includegraphics[width=1.00\textwidth]{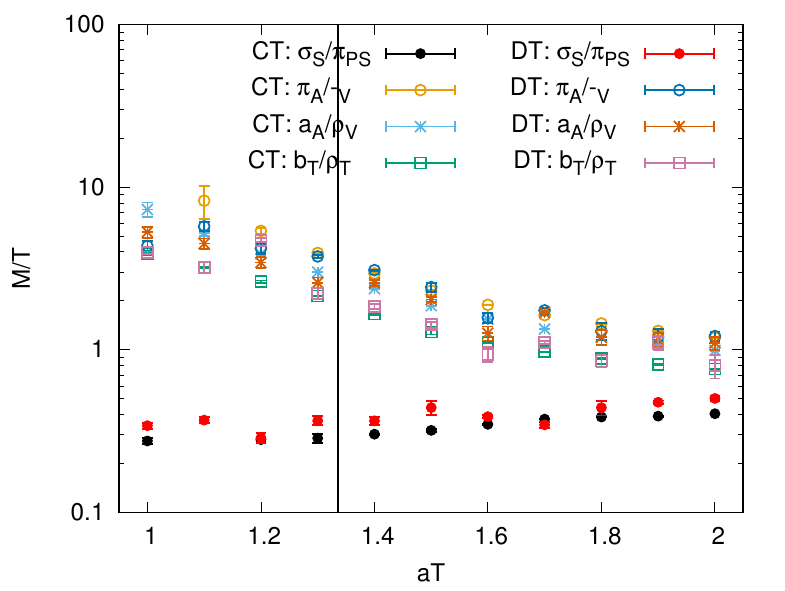}
\caption{$M/T$ for different temepratures $aT$ and channels measures either in CT or DT. Certain channels listed in (\ref{massDegen}) are mass degenerated. Agreement between CT and DT is visible.}
\label{massesDTCT}
\end{subfigure}
\begin{subfigure}[c]{0.04\textwidth}
\phantom{....}
\end{subfigure}
\begin{subfigure}[c]{0.45\textwidth}
\includegraphics[width=1.00\textwidth]{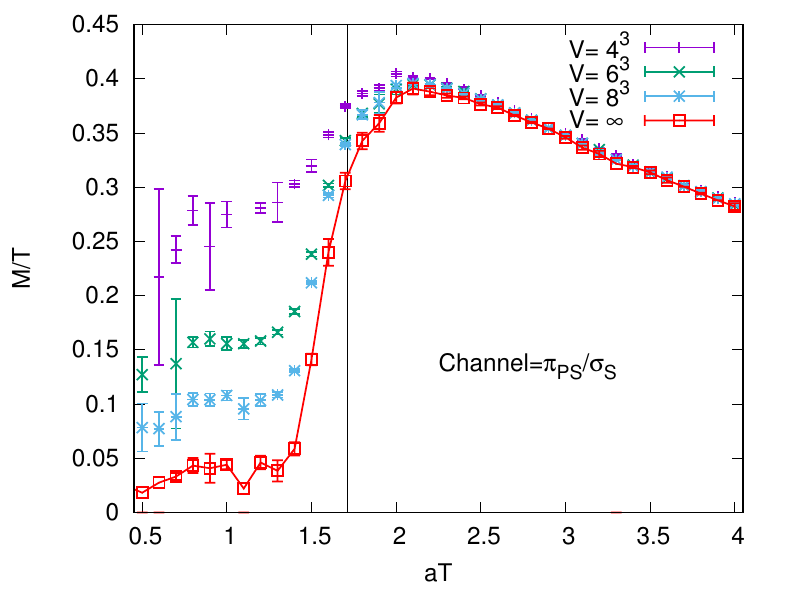}
\caption{$M/T$ over $aT$ for different volumes but the same channel $\pi_{PS}/\sigma_S$. The continuum extrapolated red curve supports the expectations to have a zero mass below $T_c$ for simulations in the chiral limit.}
\label{massesCTContinuum}
\end{subfigure}
\label{mdpConfig}
\end{figure}
Finally, Fig.~(\subref{massesDTCT}) shows a comparison of the extrapolated masses. For SU(3) the chiral transition is located at $aT\approx 1.403$ where indeed an impact on the masses is obtained. So far, only a small temperature range $aT=\{1.0,\dots 2.0\}$ is studied since especially small temperatures are expensive and simulations with reliable outcome are presented in the future. Simulations performed in discrete and continuous time give consistent results. Currently, there are still larger fluctuations and errorbars present in the discrete data. On the contrary, continuous time results have a smoother behavior down to the chiral transition but then more statistics and simulation time is needed. For $N_f=1$ we find a mass degeneracy for the channel pairs:
\begin{align}
\sigma_S \leftrightarrow \pi_{PS} \quad\quad\quad
\pi_A \leftrightarrow -_V \quad\quad\quad
b_T(\gamma_j\gamma_k) \leftrightarrow \rho_T(\gamma_i\gamma_0) \quad\quad\quad
a_A(\gamma_i\gamma_5) \leftrightarrow \rho_V(\gamma_i).
\label{massDegen}
\end{align}
Since simulations are performed in the chiral limit in a finite volume ($\epsilon-$-regime) the mass degeneracy of the Non-Oscillating scalar channel with the Oscillating pseudo-scalar channel is expected.
Due to finite volume effects non-zero $M/T$ values are obtained also in the regime below $T_c$, however, a first continuum extrapolation in Fig.~(\subref{massesCTContinuum}) clearly corrects this towards $M/T\rightarrow 0$. %shows that masses are extrapolating towards zero for temperatures below $T_c$ and then show a sudden rise but decay again. 
\section{Conclusion}
With the CT worm algorithm we measured monomer-monomer two-point correlation functions and constructed temporal correlators with projected zero spatial momentum. For a temperature range around the chiral transition we obtain consistent pole mass results for discrete and continuous time simulations. Due to simulations being tremendously more expensive for small temperature further analysis is in progress. 

The zero momentum meson correlators can be used to calculate the diffusion constant by extracting the spectral function from the correlation data applying standard methods like MEM. In continuous time we profit from being able to choose the temporal discretization by orders finer compared to discrete time computations.
%Typically, this procedure is hindered by the number of available data points in temporal direction being of $\mathcal{O}(10^1)$ while it is required to have points of order $\mathcal{O}(10^3)$. Thus, in continuous time we profit from being able to choose the temporal discretization by orders finer compared to discrete time computations.

Future calculations will be performed for finite quark masses, multiple flavors by making use of a Hamiltonian formulation \cite{unger2012} which will control and remove the sign problem and finally by including $\beta$-corrections to move away from the strong coupling limit.
%So far calculations were only performed in the chiral limit, however, there are no algorithmic limitations to not go to finite quark masses. Furthermore, even though a severe sign problem is introduced already in the mesonic sector for $N_f=2$ simluations, making use of a Hamiltonian formulation will control the sign problem and remove it. Finally, we are interested in introducing $\beta$-corrections in order to go away from the strong coupling limit towards continuum QCD.
\section{Acknowledgments}
Numerical simulations were performed on the OCuLUS cluster at PC2 (Universität Paderborn). We acknowledge support by the Deutsche Forschungsgemeinschaft (DFG) through the Emmy Noether Program under Grant No. UN 370/1 and through the Grant No. CRC-TR 211 “Strong-interaction matter under extreme conditions”.

\end{document}